# Management of Multiple Mobility Protocols and Tools in Dynamically Configurable Networks

Jochen Eisl[1], Michael Georgiades[2], Tony Jokikyyny[3], Roksana Boreli[4], Eranga Perera[4], and Kostas Pentikousis[5]

[1]Nokia Siemens Networks Germany, St.-Martin-Str. 76, D-81541 München, Germany.
Tel.: +49 89 636 75125. Fax: Fax: +49 89 636 44733  E-mail: jochen.eisl@nsn.com
[2] Department of Electrical Engineering, University of Surrey, UK.  E-mail: m.georgiades@ieee.org
[3]Ericsson, Finland.  E-mail: tony.jokikyyny@ericsson.com
[4] [4]National ICT Australia, Sydney, Australia.  Email: firstname.lastname@nicta.com.au
[5]VTT Technical Research Centre of Finland, Kaitoväylä 1, FI-90571 Oulu, Finland.  E-mail: kostas.pentikousis@vtt.fi

*Abstract*— Solutions for mobility management in wireless networks have been investigated and proposed in various research projects and standardization bodies.  With the continuing deployment of different access networks, the wider range of applications tailored for a mobile environment, and a larger diversity of wireless end systems, it emerged that a single mobility protocol (such as Mobile IP) is not sufficient to handle the different requirements adequately.  Thus a solution is needed to manage multiple mobility protocols in end systems and network nodes, to detect and select the required protocols, versions and optional features, and enable control on running daemons.  For this purpose a mobility toolbox has been developed as part of the EU funded Ambient Networks project. This paper describes this modular management approach and illustrates the additional benefits a mobility protocol can gain by using state transfer as an example.

## I. Introduction

Over the last two decades we witnessed the "unwiring" of network connectivity.  The increasing popularity of wireless services and devices has relieved users from the need for wired network access.  First, it was analog cellular voice communication.  Then, the introduction of 2G/GSM networks supported digital voice communication, limited packet data rates, and made an important step towards achieving global roaming and reachability.  3G/UMTS mobile networks extended this notion of mobility to data services with access to Internet services as the most prominent example.  These services include support for roaming (i.e. provide an identifier, where a mobile node is always reachable from any other host) and connectivity (i.e. applications running on the device can continue uninterrupted despite changes in the point of attachment.  Soon, with the growing penetration of other wireless technologies, such as IEEE 802.11 (WiFi) and IEEE 802.16 (WiMax), multiaccess and handovers between heterogeneous types of wireless access networks will become a central feature of the forthcoming telecommunications environment.  It is safe to assume that, in the near- to mid-term future, IP connectivity will increase in importance and the so-called "fixed mobile convergence" will be based on the TCP/IP protocol suite, into which new functions will have to be integrated.

Thus, a starting point for building new solutions has to be the Mobile IP protocol, which has been the most prominent mobility protocol for several years [1][2][3].  In addition, approaches have been taken to provide mobility solutions at layers other than the network layer, for example, at the transport layer [4] or above it [5][6].  Another proposal is to add a layer between transport and network layers in order to differentiate between (and disentangle) the "host identity" from its "location" [7].  But mobility solutions abound at the network layer as well.  After a number of shortcomings were identified with the original Mobile IP proposal, including non-localized location management, the impact on packet delivery due to triangular routing, and packet loss during handovers.  Hence, the concept of micro-mobility was introduced, with HMIP [8], Cellular IP [9], HAWAII [10], and NETLMM [11] among the prominent examples in this area of research and development.  Moreover, several techniques and mechanisms were also proposed for enabling seamless mobility [12][13][14].

Ambient Networks project (www.ambient-networks.org) [15][16] aims at defining and developing solutions for a future wireless world which will be filled by a multitude of user devices and wireless technologies and network access will be affordable, simple-to-use, and available anytime-anywhere.  This vision leads to a diverse set of requirements that none of the existing mobility management proposals can address. We argue that the consequence of the cornucopia of mobility management solutions is that it is no longer possible to envisage a single mobility management paradigm.  In previous work we introduced the concept of a set of mobility solutions that can be flexibly combined to support specific mobility events – effectively providing a "mobility toolbox" [17][18].  In this paper, we overview our modular management approach and illustrate the additional benefits a mobility protocol or tool can gain.  In particular, we use state transfer as an example of protocol integration into the toolbox.

The following section presents the framework for the mobility toolbox, like it has been suggested for the EU project Ambient Networks (AN).  To provide relevant examples for mobility protocol selection, we outline the several scenarios and explain the interactions between end nodes and network nodes in order to accomplish the selection task. In Section III we focus on a mobility support protocol, namely state transfer, to illustrate the additional advantages/benefits that the mobility toolbox has to offer. In Section IV we briefly present our ongoing effort in developing the Ambient Networks prototype. The state transfer module is currently realized and, in a next step, the mobility toolbox will be augmented with a state transfer protocol. We provide an overview of the design and implementation for these functions before we conclude the paper in Section V.

## II. AN OVERVIEW OF THE MOBILITY TOOLBOX

One of the main challenges in today's networking world is how to address mobility, heterogeneity and the integration of networks without having to plug-and-play various services in an *ad-hoc manner*. In order to answer this question the AN architecture has introduced the novel concept of the Ambient Control Space (ACS) [15][16]. The ACS is an overlay network control layer that brings the control functions for heterogeneous networks under one umbrella, and facilitates functional decomposition of the different control elements into distinct functional entities (FEs). The centerpiece in our design is the *Handover and Locator Management* FE (HOLM) comprising protocols and mechanisms for handling mobility management as well as being a coordinating module for tool selection [18]. This coordinating module together with the mobility protocols are referred to as the "mobility toolbox" in our architecture. HOLM integrates a set of protocols, some of which are included into the ACS with little or no modification, such as Mobile IP [19] and HIP [7]. The toolbox interacts with these protocols via a well-defined interface (e.g. BSD sockets).

It should be stressed that "HOLM" refers to a *functional entity*, not a particular implementation. HOLM is not a monolithic set of protocols or modules, which is available at every node with all proposed functionality, but instead the appropriate modules are used based on specific node requirements. For example, a handheld device, a laptop and a core network node will most likely feature a different set of modules and protocols, as illustrated in Fig. 1, where four different instantiations of the mobility toolbox are implemented at different hosts as part of the overall control space (titled ACS 1 to 4).

The following subsections present specific scenarios where it is not straightforward which protocol (version and feature(s)) should be used. This is due to spontaneous requirements (for instance by applications) or available network capabilities (in a candidate network to be visited). As one of the anonymous reviewers of this paper commented, "although most mobility protocols and extensions include compatibility detection, relying on such techniques alone would result in potentially long trial-and-error procedures during which a mobile host would attempt to discover protocol support in the network and at the correspondent host. This approach also limits the correspondent host and the network in influencing protocol selection." We argue that with HOLM in place, future mobile hosts will not run into the deadlocks or suboptimal tool and protocol selection as described in the following scenarios for mobility tool selection. The scenarios described next are illustrated in Fig. 1.

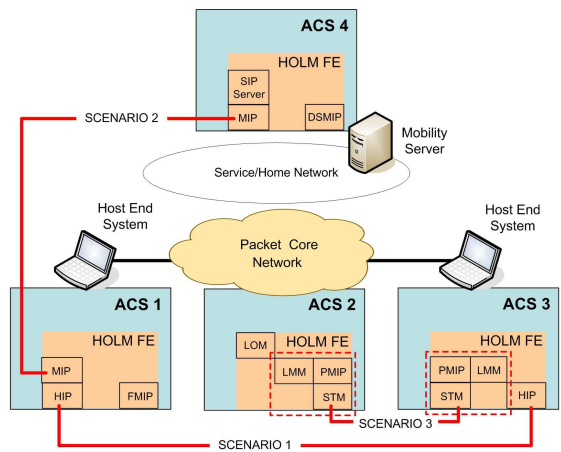

Fig. 1. Handover tool selection scenarios

### A. Scenario 1: End-Host Constraints and Tool Selection

In this case a mobile node negotiates and agrees with a correspondent node (possibly mobile too) about the mobility protocol(s) to be used between the two peers. The primary consideration is the supported *protocol* with the appropriate version for both peers. If both parties have more than one option for mobility management built-in, peer-to-peer communication may introduce further constraints for the selection of the appropriate protocol. Hosts may agree to use a protocol based on specific criteria such as user / operator preferences and policies, protocol stack(s) (IPv4 vs. IPv6), and additional supported features, such as security support, multi-homing support, supported applications and handover performance metrics. The following list of examples where selection between candidate tools and protocols is necessary is not claimed to be complete:

- Protocol stack: MIPv4 should be chosen because one of the hosts does not support IPv6, and the access gateway provides a service tunneling IPv6 networks.
- Security: HIP because it uses security associations built into peer-to-peer connections.
- Supported applications: TCP (migrate) [20] can be used for a large variety of applications (not just those based on SIP) when there are no specific requirements from the applications
- Application preferences: SIP is used because the host runs peer-to-peer applications using SIP.
- Multihoming support: SCTP [21] can support multiple sessions at the transport layer.
- Performance considerations: MIPv6 with route optimization because it decreases the overall round trip time when compared to standard MIPv6.

*B. Scenario 2: Service Provider Network Tool Selection*

As opposed to the previous scenario, the aspects addressed here are not related with which protocol is used, but what *feature* of a certain protocol is used. The most obvious example for this scenario is related to communication between a MIPv4/v6 mobile node and its home agent. In this use case, a service provider network may implement the MIP home agent and a variety of add-on features, which could be requested by the client network and/or node. Examples of such protocol features include (i) protecting MIPv6 messages with IPSec [22], (ii) IPv6 support for firewall traversal [23], (iii) support of dual stack operation (MIPv4 and MIPv6) [24], (iv) support for home agent reliability [25], (v) multiple care-of address registration [26], (vi) support for network mobility [27], and (vii) Mobile IPv4 dynamic home agent assignment [28].

In principle, this scenario is not limited to MIP home agents but can be applied to other middle boxes, which provide support for other mobility protocols.

*C. Scenario 3: Access Network Provider Tool Selection*

In this use case a client host and/or network selects specific tools from the access provider network. Some mobility services are constrained to the local access. There are several aspects how tool selection could work between a mobile client and an access network. First, a mobile host needs to transfer state from the current access to the new access network (e.g. for header compression) and therefore the state transfer service needs to be activated. Second, in order to support seamless handover, different techniques may be available. FMIPv6, for example, establishes the new network path *before* the old one is disconnected and provides a forwarding mechanism of packets from the old to the new access in order to avoid packet loss [12].

Moreover, a legacy node would require support from a proxy node for mobility management [29]. In that case, the proxy node generates the mobility related signaling on behalf of the mobile client. A further example is a client node selecting MIPv4 because a Foreign Agent is detected. Finally, support for common radio resource management can also be considered in this scenario. Mobile node(s) and access networks may support a common radio control layer, such as the Media Independent Handover Services currently under standardization by IEEE 802.21 [30], to exchange radio link-layer notifications and offer command services for the common handling of radio resources.

More information on the Ambient Network Mobility Scenarios & Requirements are available in [31].

## III. MOBILITY TOOLBOX AND STATE TRANSFER

In this section we focus on the scenario described is Section II.C above, and use the State Transfer Module (STM) as an example of a tool that can be added into HOLM and the ACS. First, in the following subsection, we describe the additional benefits that STM (and any other tool for that matter) can gain by being integrated into HOLM. Then we present its operation in subsection III.B.

*A. State Transfer Module—An Overview*

The Context Transfer Protocol (CTP) [14] aims at minimizing the impact of certain transport, routing, and security-related services on handover performance. When a mobile node moves to a new subnet it needs to continue using the services that have already been established while connected at the previous subnet. Within the HOLM framework, we propose STM which is designed to forward state transfer information related to the mobile host's sessions from a previous access router to a next access router when handoff takes place in a mobile communication network. STM is set to reside at one of the access routers and to provide a message framework for the interworking between itself, the Handover Selection and Execution Control (HOSEC) module, which is presented in [32], and the state transfer candidate service protocols. STM defines a message sequence exchange framework for forwarding state information between itself and a peer STM residing at another access router.

Although the design of STM is primarily based on CTP, it is integrated with the mobility toolbox and thus provides several new features. STM can utilize triggers [33] received by the mobility tool box providing better synchronization with a plurality of mobility management protocols and also not necessarily relying on the mobile terminal to initiate transfer. State transfer can take place between any two AN nodes supporting STM and it is designed in alignment with the Ambient Networks concepts [16], offering a possible context forwarding tool to AN functional entities. Furthermore, it makes use of the AN naming scheme for identifying state transfer source and destination and uses the AN security scheme for obtaining security associations between the involved nodes prior to state transfer. For signaling, STM makes use of the generic signaling layer protocol (GSLP) proposed in AN for transferring the state and it is designed to utilize the generic transport layer protocol (GTLP) proposed in AN [31][34]. STM can also work over UDP and TCP.

*B. State Transfer Module—Operation*

STM requires the implementation of different components and interfaces to interact with other modules of HOLM and the ACS. It consists of a module as part of the HOLM FE which contains the logic/state machine for the state transfer operation. This is used for handling state information at the initiator as well as installing state information at the receiver. STM also defines:

- An interface for receiving triggers
- An API to allow other functional entities to make use of STM. This interface can be used by other STM candidate services (e.g. QoS, Header Compression, AAA)
- A protocol to carry the information and signaling between the STMs of the involved ACSs.

Fig. 2 illustrates an example of STM in operation requesting firewall support as it performs a handover. In steps (1) and (2) the mobile node MN initiates two RTP [35] sessions with corresponding nodes CN1 and CN2, respectively. In step (3), the mobile node MN handoffs from its radio access network (RAN1) to another one (RAN2). At step (4), during the handover, the Triggering FE [33] at the mobile node sends a trigger to the STM at firewall FW2 with the RAN1 routing/address information. In step (5), the STM at FW2 sends a Request Message to STM at firewall FW1, and in (6) the STM at firewall FW1 replies with an "STM Res Message". The mobile node MN is now in a position to send message (7), a SIP Re-Invite to CN1, (8) a SIP Re-Invite to CN 2 and (9) a SIP Register (MN) to the SIP server [6].

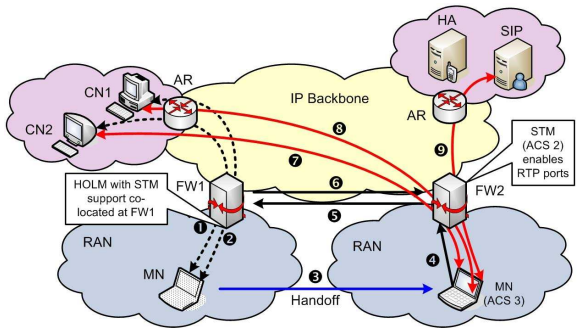

Fig. 2. STM firewall support

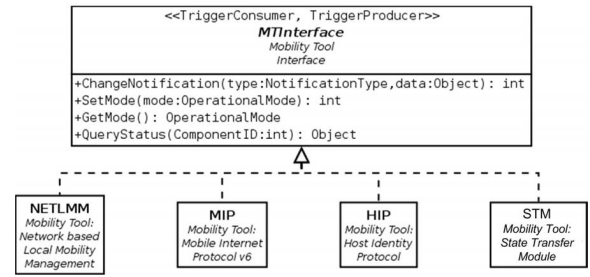

Fig. 3. Mobility tool interface

Once the STM at FW2 receives the "STM Res Message" it parses it to check the context payload. It will then be in a position to configure the firewall according to the received context. In this scenario, it enables specific RTP ports associated with the mobile node's sessions to avoid any session interruption or traffic blocking.

## IV. PROTOTYPE IMPLEMENTATION

The project develops an integrated prototype [36] to demonstrate and verify the feasibility of the most prominent functions of Ambient Networks. In this section we describe the design for the mobility toolbox and the state transfer module. These are currently implemented and in a next step the will be integrated into the common prototype.

First, the tools register themselves with the toolbox module and implement the mobility tool interface (MTI). The communication between different entities can be implemented using TRG [33]: Each tool registers to listen to "MTI-common" triggers types, and specialized "MTI-xxx", where xxx is replaced by the tool's name. The MTI-common "channel" is used to provide triggers in a broadcast manner, whereas the MTI-xxx triggers are used for private communication between the toolbox and a specific tool.

### A. Mobility Tool Interface

Fig. 3 illustrates the Mobility Tool Interface. The essential interface functions in addition to the ordinary register and deregister functions are defined as follows.

*Change Notifications*—This is a bi-directional interface between the toolbox and a protocol module, say the HIP [7] implementation. Triggers are sent by the toolbox to notify the protocol module about external events, such as changes in locators, interfaces, locator domains, and so on, just to name a few, allowing the protocol(s) to act accordingly. Conversely, the interface can be used by a protocol module to inform the toolbox about any changes in the protocol state machine, completed actions, etc. It is still for further investigation whether notification types have to be individually defined for each protocol. In terms of implementation, it is assumed that the data transferred is specific for each protocol but can be derived from and/or inherited by an abstract class (in object-oriented programming terms).

*Set Mode*—The toolbox uses this message to set a specific protocol into certain mode or query for information. Valid modes are presented in subsection IV.C. This operation is needed to control a protocol state machine, for instance, to prevent a protocol state machine to be changed by certain events, such as, inbound Mobile IP "Router Advertisement" messages .

*Get Mode*—This function is used by the toolbox to obtain the current operational mode for a particular protocol daemon.

*Query Status*—The toolbox asks the status of a protocol's specific information element(s), for instance, "how many hosts are attached to a certain mobile network". This information is protocol-specific.

Next we introduce the details of the STM-MTI interface of Fig. 3 describing the interaction between toolbox and STM.

### B. MTI Notification Types

The following primitives are related to the MTI "Change Notification" function.

STM PUSH—STM receives a request from the toolbox to initiate state transfer to certain STM peer(s). This message includes the list of possible STM types that could be transferred.

STM PULL—STM receives a request from the toolbox to initiate state transfer from a certain STM peer that is to pull state to a new location from the location currently maintaining it. This message includes a list of the possible STM types that should be transferred.

STM ACCEPT—This notification is sent from STM to the mobility toolbox so that the STM peer(s) agree to the state transfer request, an indication on which state could or could not be sent and an ID for the transfer.

STM START—This message initiates state transfer after it has been accepted by the STM peer. The mobility toolbox may reject the state transfer, if the required STM type cannot be supported by the peer STM.

STM ACK—Used for acknowledging a successful state transfer operation between the STM peers to the toolbox. This message must be sent immediately after the peer STM has acknowledged the installation of the state.

STM ERROR—This carries an error message indicating to the toolbox that either the source or target STMs would like to request retransmission or terminate the process due to an unexpected failure.

*C. Operation Modes*

The following primitives are related to the MTI "Set Status" function presented earlier.

STM INIT—Request to initialize the STM module on a node with the necessary configuration parameters, for example, TCP port number for communication, timer values, and so on.

STM ABORT—It has been recognized by the toolbox that the current state transfer is invalid for some reason and instructs the initiating STM peer to abort the operation.

STM PAUSE—In some cases, the state transfer operation can be interrupted, for example, if the resources are needed for some other state transfer with higher priority.

STM CONT—A transfer, which has come to a halt may be continued later on, if the necessary requirements are met. This is akin to continuing an interrupted FTP session.

STM TERMINATE—The STM daemon on the local node (co-located with the mobility toolbox) will stop its operation and will be terminated. This is especially needed for system shutdowns.

## V. CONCLUSION AND FUTURE WORK

We motivated the need for the co-existence of multiple mobility management protocols and tools in devices targeting heterogeneous networks, which provide convergence among different types of accesses and network services. Further we expect that networks and devices need to cope with a variety of different protocols and new features and so far it has been unclear how this selection process can be handled dynamically in a self-configured way. Our proposed solution is a mobility toolbox, which has been designed and is currently implemented within the Ambient Networks project.

This paper presented tool selection scenarios where the value of our mobility toolbox can be seen. We then considered state transfer as an example of a mobility support protocol that can be integrated with the toolbox and discussed the benefits that state transfer can reap from the availability of the mobility toolbox. We defined an interface between the toolbox and the individual protocol module and provided a specification of how the interaction between the state transfer implementation and the mobility toolbox should work.

We are currently proceeding with completing the component implementations, in particular the state transfer module, and we will shortly start the final integration of all toolbox-related modules into the common Ambient Networks prototype. As we proceed demonstrating the feasibility of the toolbox concept and the state transfer module, we expect to measure significant gains in handover performance based on the early detection of the supported feature sets for mobility protocol. Further gains are expected from controlling the operational state of different mobility management protocols and tools.

ACKNOWLEDGMENT

This work has been carried out in the framework of the Ambient Networks project (IST 027662), which is partially funded by the Commission of the European Union. The views expressed in this paper are solely those of the authors and do not necessarily represent the views of their employers, the Ambient Networks project, or the Commission of the European Union. The ideas from people involved in the projects and mobility management research are gratefully acknowledged. So are their comments and suggestions.

Last, but certainly not least, we would like to thank the anonymous reviewers of the First Ambient Networks Workshop on Mobility, Multiaccess, and Network Management (M2NM). Their comments and suggestions have improved this paper considerably.

REFERENCES

[1] C. Perkins, *Mobile IP Design Principles and Practices*, Addison-Wesley, 1998.
[2] C.E. Perkins, "Mobile IP", *IEEE Communications Magazine*, vol.40, no.5, pp.66-82, May 2002.
[3] D. Johnson, C. Perkins, and J. Arkko, *Mobility Support in IPv6*, IETF RFC 3775, June 2004.
[4] H. Izumikawa, I. Yamaguchi, and J. Katto, "An efficient TCP with explicit handover notification for mobile networks*", Proc. IEEE Wireless Communications and Networking Conference* (WCNC), Atlanta, Georgia, USA, March 2004 pp. 647-652.
[5] E. Wedlund and H. Schulzrinne, "Mobility support using SIP", *Proc. Second ACM International Workshop on Wireless Mobile Multimedia*, Seattle, Washington, USA, August 1999.
[6] J. Rosenberg, H. Schulzrinne, G. Camarillo, A. Johnston, J. Peterson, R. Sparks, M. Handley, and E. Schooler, *SIP: Session Initiation Protocol*, , IETF RFC 3261, June 2002.
[7] R.Moskowitz and P.Nikander, *Host Identity Protocol (HIP) Architecture*, IETF RFC 4423, May 2006.
[8] H. Soliman, C. Castelluccia, K. El Malki, and L. Bellier, *Hierarchical Mobile IPv6 Mobility Management (HMIPv6)*, IETF RFC 4140, August 2005.
[9] A. T. Campbell, J. Gomez, A. G. Valkó, "An overview of cellular IP", *Proc. IEEE Wireless Communications and Networking Conference* (WCNC), September 1999 pp. 606-610.
[10] R. Ramjee, K. Varadhan, L. Salgarelli, S. R. Thuel, S.-Y. Wang, and T. La Porta, "HAWAII: A domain-based approach for supporting mobility in wide-area wireless networks", *IEEE/ACM Transactions on Networking*, vol. 10, no. 3, June 2002 pp. 396-410.
[11] J. Kempf (Ed.), *Goals for Network-based Localized Mobility Management (NETLMM)*, IETF RFC 4831, April 2007.
[12] R. Koodli (Ed.), *Fast Handovers for Mobile IPv6*, IETF RFC 4068, July 2005.

[13] M. Liebsch (Ed.), A. Singh (Ed.), H. Chaskar, D. Funato, and E. Shim, *Candidate Access Router Discovery (CARD)*, IETF RFC 4066, July 2005.

[14] J. Loughney (Ed.), M. Nakhjiri, C. Perkins, and R. Koodli, *Context Transfer Protocol (CXTP)*, IETF RFC 4067, July 2005.

[15] N. Niebert, A. Schieder, H. Abramowicz, G. Malmgren, J. Sachs, U. Horn, C. Prehofer, and H. Karl, "Ambient Networks – An architecture for communication networks beyond 3G", *IEEE Wireless Communications*, vol. 11, no. 2, pp. 14–21, April 2004.

[16] N. Niebert, A. Schieder, J. Zander, and R. Hancock (Eds.), *Ambient Networks, Co-operative Mobile Networking for the Wireless World*, Wiley, April 2007.

[17] T. Toniatti, F. Meago, A. Periccioli, S. Uno, E. Perera, and R. Boreli, "Advanced Handover Management in Ambient Networks", *Proc. International Workshop on Convergent Technologies* (IWCT), Oulu, Finland, June 2005.

[18] R. Aguero, A. Surtees, J. Eisl, and M. Georgiades, "Mobility Management in Ambient Networks", *Proc. 65$^{th}$ IEEE Vehicular Technology Conference* (VTC 2007-Spring), Dublin, Ireland, April 2007, pp.894-898.

[19] C. Perkins (Ed.), *IP Mobility Support for IPv4*, IETF RFC 3344, August 2002.

[20] A. C. Snoeren and H. Balakrishnan, "An End-to-End Approach to Host Mobility", Proc. 6th ACM MobiCom, August 2000.

[21] M. Riegel, M. Tüxen, N. Rozic Nikola, and D. Begusic Dinko, "Mobile SCTP transport layer mobility management for the Internet", *Proc. International Conference on Software, Telecommunications and Computer Networks* (SoftCOM), Split, Croatia, October 2002.

[22] J. Arkko, V. Devarapalli, and F. Dupont *Protecting MIPv6 messages with IPSec*, IETF RFC 3776, July 2005.

[23] F. Le, S. Faccin, B. Patil, and H. Tschofenig, *Mobile IPv6 and Firewalls: Problem Statement*, IETF RFC 4487, May 2006.

[24] H.Soliman (Ed.), *Mobile IPv6 support for dual stack Hosts and Routers (DSMIPv6)*, IETF Internet Draft, July 2007. Work in progress.

[25] R. Wakikawa (Ed.), *Home Agent Reliability Protocol*, IETF Internet Draft, July 2007. Work in progress.

[26] R. Wakikawa, T. Ernst, K. Nagami, and V. Devarapalli, *Multiple Care-of Addresses Registration*, IETF Internet Draft, July 2007. Work in progress.

[27] V. Devarapalli, R. Wakikawa, A. Petrescu, and P. Thubert, *NEtwork MObility (NEMO) Basic Support Protocol*, IETF RFC 3963, January 2005.

[28] M. Kulkarni, A. Patel, and K. Leung,*Mobile IPv4 Dynamic Home Agent (HA) Assignment*, IETF RFC 4433, March 2006.

[29] S. Gundavelli, K. Leung, V. Devarapalli, K. Chowdury, B. Patil, *Proxy Mobile IPv6*, IETF Internet Draft, June 2007. Work in progress.

[30] A. Rajkumar, M. Williams, X. Liu, G. Vivek (Eds.), *Draft IEEE Standard for Local and Metropolitan Area Networks: Media Independent Handover Services*, IEEE P802.21/D02.00, September 2006. Work in progress.

[31] J. Eisl (Ed.), *Ambient Network Mobility Scenarios & Requirements*, WWI Ambient Networks, Public Deliverable D4-1, July 2005.

[32] H. Tang and J. Eisl (Eds.), *Mobility Support: Design and Specification*, Ambient Networks Phase 2, Public Deliverable D09-B1, December 2006.

[33] J. Mäkelä and K. Pentikousis, "Trigger management mechanisms", *Proc. Second International Symposium on Wireless Pervasive Computing* (ISWPC), San Juan, Puerto Rico, USA, February 2007.

[34] N. Akhtar, R. Campos, C. Kappler, P. Paakkonen, P. Poyhonen, D. Zhou, "GANS: A Signalling Framework for Dynamic Interworking Between Heterogeneous Networks", *Proc. 64th IEEE Vehicular Technology Conference* (VTC 2006-Fall), Montreal, Canada, September 2006, pp.1-5.

[35] H. Schulzrinne, S. Casner, R. Frederick, and V. Jacobson, *RTP: A Transport Protocol for Real-Time Applications*, IETF RFC 1889, January 1996.

[36] P. Pääkkönen, P. Salmela, R. Aguero, and J. Choque, "An integrated Ambient Networks prototype", *Proc. International Conference on Software, Telecommunications and Computer Networks* (SoftCOM), Split, Croatia, September 2007.